\documentclass[10pt]{article}
\usepackage[dvips]{graphicx}
\begin{document}
\begin{center}
\textbf{{\large  Potentials of Coupled Quintessence Based on a
Dilaton}}
 \vskip 0.35 in
\begin{minipage}{4.5 in}
\begin{center}
{\small Z. G. HUANG$^{1,~\dag}$, Q. Q. SUN$^{1}$ \vskip 0.06 in
\textit{ $^1$Department~of~Mathematics~and~Physics,
\\~Huaihai~Institute~of~Technology,~222005,~Lianyungang,~China
\\
$^\dag$zghuang@hhit.edu.cn}} \vskip 0.25 in {\small W. FANG$^{2}$
and H. Q. LU$^{2,~\ddag}$ \vskip 0.06 in \textit{
$^2$Department~of~Physics,~Shanghai~University,~Shanghai,~China
\\
$^\ddag$alberthq$\_$lu@staff.shu.edu.cn}}
\end{center}
\vskip 0.2 in

{\small In this paper, we regard dilaton in Weyl-scaled induced
gravitational theory as coupled Quintessence, which is called DCQ
model by us. Parametrization of the dark energy model is a good
method by which we can construct the scalar potential directly from
the effective equation of state function $\omega_\sigma(z)$
describing the properties of the dark energy. Applying this method
to the DCQ model, we consider four parametrizations of $\omega(z)$
and investigate the features of the constructed DCQ potentials,
which possess two different evolutive behaviors called $"O"$ mode
and $"E"$ mode. Lastly, we comprise the results of the constructed
DCQ model with those of quintessence model numerically.\vskip 0.2 in
\textit{Keywords:} Dark energy; Dilaton; Coupled Quintessence;
Parametrization; Potential.
}
\end{minipage}
\end{center}
\vskip 0.2 in
\begin{flushleft}\textbf{1. Introduction}\end{flushleft}

Data of high-redshift Type Ia Supernova[1] and the Cosmic Microwave
Background[2] have shown us such a fact: the density of clustered
matter including cold dark matters plus baryons,
$\Omega_{m0}\sim1/3$, and that the Universe is flat to high
precision, $\Omega_{total}=0.99\pm0.03$[3]. That is to say, we are
living in a flat universe which is undergoing a phase of accelerated
expansion, and there exists an unclumped form of energy density
pervading the Universe. This unknown energy density which is called
"dark energy" with negative pressure, contributes to two thirds of
the total energy density. To know the nature of dark energy is one
of the most challengeable problems in cosmological research. So far,
theoretical physicists and astrophysicists have constructed many
models including cosmological constant $\Lambda$,
Quintessence[4-14], K-essence[15], Tachyon[16], Phantom[17-20],
Quintom[21] and holographic dark energy[22] so on, to fit these
observations. Perhaps the simplest explanation for these data is
that the dark energy corresponds to a positive cosmological
constant. However, the cosmological constant model suffers from two
serious issues called "coincidence problem" and "fine-tuning
problem". The essential characteristics of these dark energy models
are contained in the parameter of its equation of state,
$p=\omega\rho$, where $p$ and $\rho$ denote the pressure and energy
density of dark energy, respectively, and $\omega$ is a state
parameter. Quintessence model has been widely studied, and its state
parameter $\omega_\phi$ which is time-dependent, is greater than
$-1$. Such a model for a broad class of potentials can give the
energy density converging to its present value for a wide set of
initial conditions in the past and posses tracker behavior. The
quintessence potential $V(\phi)$ and the equation of state
$\omega_\phi(z)$ may be reconstructed from supernova
observations[23]. Guo et al.[24] have constructed a theoretical
method of constructing the quintessence potential $V(\phi)$ directly
from the dark energy equation of state function $\omega_\phi(z)$.
They investigated the general features of quintessence potentials
and obtained that the typical behavior of the potentials is a
runaway type.
\par In this paper, we regard dilaton in Weyl-scaled induced
gravitational theory as a coupled Quintessence[25]. We call our
model DCQ model. We apply the theoretical method of parametrization
of dark energy to the DCQ model and consider four typical
parametrization of $\omega_\sigma(z)$ as follows[26-31]: Case I:
$\omega_\sigma=\omega_0$; Case II:
$\omega_\sigma=\omega_0+\omega_1z$; Case III:
$\omega_\sigma=\omega_0+\omega_1\frac{z}{1+z}$; Case IV:
$\omega_\sigma=\omega_0+\omega_1ln(1+z)$, which fit the observations
well. In these four cases, the properties of the constructed DCQ
potentials $W_\sigma(z)$ are considered and the evolutions of the
dark energy density $\rho_\sigma(z)$  with respect to $z$ are also
shown mathematically. \vskip 0.2 in
\begin{flushleft}\textbf{2. Basic Equations}\end{flushleft}
\par Now let us consider the action of the Weyl-scaled induced gravitational theory:
\begin{equation}S=\int{d^4X\sqrt{-g}[\frac{1}{2}R(g_{\mu\nu})-\frac{1}{2}g^{\mu\nu}\partial_\mu\sigma\partial_\nu\sigma-W(\sigma)+L_{fluid}(\psi)}]\end{equation}
where the lagrangian density of cosmic fluid
$L_{fluid}(\psi)=\frac{1}{2}g^{\mu\nu}e^{-\alpha\sigma}\partial_\mu\psi\partial_\nu\psi-e^{-2\alpha\sigma}V(\psi)$,
$\alpha=\sqrt{\frac{\kappa^2}{2\varpi+3}}$ with $\varpi>3500$[32]
being an important parameter in Weyl-scaled induced gravitational
theory, $\sigma$ is DCQ field, $W(\sigma)$ is DCQ potential,
$g_{\mu\nu}$ is the Pauli metric which can really represent the
massless spin-two graviton and should be considered to be physical
metric[33]. This means that, as long as one wants to interpret the
theory as a generalization of Einstein' theory, one must treat the
Pauli metric as physical. In the Kaluza-klein theory Cho also
arrives essentially at the same conclusion, but for a different
reason. One cannot treat the Jordan metric as physical, because it
violates the positivity of the Hamiltonian. At the same time, one
must accept the Pauli metric as physical, as long as one wishes to
achieve the unification of Einstein's gravitation with other
interactions from the Kaluza-Klein theory. We work in
units($\kappa^2\equiv8\pi G=1$). From the solar system tests, the
current constrain is $\alpha^2<0.001$[34]. The new constrain on the
parameter is $\alpha^2<0.0001$[35], which seems to argue against the
existence of long-range scalars. Perhaps such a pessimistic
interpretation of the limit is premature [33,34]. The conventional
Einstein gravity limit occurs as $\sigma\rightarrow 0$ for an
arbitrary $\varpi$ or $\varpi\rightarrow\infty$ with an arbitrary
$\sigma$. When $W(\sigma)=0$, it will result in the
Einstein-Brans-Dicke theory.
\par By varying action(1) and working in FRW universe, we obtain the field
equations of Weyl-scaled induced gravitational theory:
\begin{equation}H^2=\frac{1}{3}[\rho_\sigma+e^{-\alpha\sigma}\rho]\end{equation}
\begin{equation}\frac{\ddot{a}}{a}=-\frac{1}{6}(e^{-\alpha\sigma}\rho+\rho_\sigma+p_\sigma)\end{equation}
\begin{equation}\ddot{\sigma}+3H\dot{\sigma}+\frac{dW(\sigma)}{d\sigma}=\frac{1}{2}\alpha e^{-\alpha\sigma}(\rho-3p)\end{equation}
\begin{equation}\dot{\rho}+3H(\rho+p)=\frac{1}{2}\alpha\dot{\sigma}(\rho+3p)\end{equation}
where $H=\frac{\dot{a}}{a}$ is Hubble parameter, and $\rho$ includes
matter energy density $\rho_m$ and the radiation energy density
$\rho_r$. In what follows, we shall neglect the radiation energy
density $\rho_r$. For matter $p_m=0$, we get $\rho_m=\rho_{m_0}
\frac{e^{\frac{1}{2}\alpha\sigma}}{a^3}$ from Eq.(5). The effective
energy density $\rho_\sigma$ and the effective pressure $p_\sigma$
of DCQ field can be expressed as follows
\begin{equation}\rho_{\sigma}=\frac{1}{2}\dot{\sigma}^2+W(\sigma)\end{equation}
\begin{equation}p_{\sigma}=\frac{1}{2}\dot{\sigma}^2-W(\sigma)\end{equation}
According to the above results, we have
\begin{equation}W(\sigma)=3H^2+\dot{H}-\frac{1}{2}\rho_{m_0}e^{-\frac{1}{2}\alpha\sigma}a^{-3}\end{equation}
\begin{equation}\dot{\sigma}^2=-2\dot{H}-\rho_{m_0}e^{-\frac{1}{2}\alpha\sigma}a^{-3}\end{equation}
The relationship between scale factor $a$ and redshift $z$ is
$z+1=\frac{a_0}{a}$ where $a_0$ denotes the value of scale factor at
redshift $z=0$(present). So, we have
\begin{equation}\dot{H}=\frac{dH}{dt}=\frac{\frac{a_0}{z+1}H}{-\frac{a_0}{(z+1)^2}}\frac{dH}{dz}=-H(z+1)\frac{dH}{dz}\end{equation}
\begin{equation}\dot{\sigma}=\frac{d\sigma}{dt}=-H(z+1)\frac{d\sigma}{dz}\end{equation}
Eqs.(8)(9) can be rewritten as
\begin{equation}W(z)=3H^2-H(z+1)\frac{dH}{dz}-\frac{1}{2}\rho_{m_0}e^{-\frac{1}{2}\alpha\sigma}(1+z)^{3}\end{equation}
\begin{equation}\dot{\sigma}^2=\frac{2}{H(z+1)}\frac{dH}{dz}-\frac{\rho_{m_0}e^{-\frac{1}{2}\alpha\sigma}(1+z)}{H^2}\end{equation}
We define the dimensionless dark energy function $\xi(z)$ as follows
\begin{equation}\xi(z)=\frac{\rho_\sigma}{\rho_{\sigma_0}}\end{equation}
where $\rho_{\sigma_0}$ is dark energy density at redshift
$z=0$(present). Using Eqs.(2)(14), we get
\begin{equation}\frac{dH}{dz}=\frac{\rho_{\sigma_0}}{6H}\frac{d\xi}{dz}+\frac{\rho_{m_0}e^{-\frac{1}{2}\alpha\sigma}(1+z)^2}{2H}-\frac{\alpha\rho_{m_0}e^{-\frac{1}{2}\alpha\sigma}(1+z)^3}{12H}\frac{d\sigma}{dz}\end{equation}
Substituting Eqs.(2)(15) into Eqs.(12)(13), we have
\begin{equation}\tilde{W}(z)=(1-\Omega_{m_0})[\xi(z)-\frac{1+z}{6}\frac{d\xi}{dz}]+\frac{1}{12}\alpha\Omega_{m_0}e^{-\frac{1}{2}\alpha\sigma}(1+z)^4\frac{d\sigma}{dz}\end{equation}
\begin{equation}\frac{d\sigma}{dz}=-\frac{1}{\chi(z)}[\frac{1-\Omega_{m_0}}{1+z}\frac{d\xi}{dz}-\frac{1}{2}\alpha\Omega_{m_0}e^{-\frac{1}{2}\alpha\sigma}(1+z)^2\frac{d\sigma}{dz}]^{\frac{1}{2}}\end{equation}
where $\Omega_{m_0}\equiv\frac{\rho_{m_0}}{\rho_0}$ is the present
matter energy density, with $\rho_0=\rho_{\sigma_0}+\rho_{m_0}$
being present total energy density, $\tilde{W}(z)=W(z)/\rho_0$, and
\begin{equation}\chi(z)=[(1-\Omega_{m_0})\xi(z)+\Omega_{m_0}e^{-\frac{1}{2}\alpha\sigma}(1+z)^3]^{\frac{1}{2}}\end{equation}
is the cosmic expansion rate relative to its present value. From
Eqs.(16)(17), we can see that the DCQ model will reduce to ordinary
quintessence model as the coupled constant $\alpha\rightarrow0$.
Obviously, Eq.(17) is a nonlinear first-order differential equation
with respect to $z$ and it is difficult to find its analytic
solutions. Next we investigate the properties of the constructed DCQ
potential numerically. In DCQ model, we take the simplest dimension
dark energy function $\xi(z)=(1+z)^{3(1+\omega_\sigma)}$. So,
Eqs.(16)(17) can be rewritten as
\begin{equation}\tilde{W}(z)=\frac{1}{2}(1-\Omega_{m_0})(1-\omega_\sigma)(1+z)^{3(1+\omega_\sigma)}+\frac{1}{12}\alpha\Omega_{m_0}e^{-\frac{1}{2}\alpha\sigma}(1+z)^4\frac{d\sigma}{dz}\end{equation}
\begin{equation}\frac{d\sigma}{dz}=-\frac{1}{\chi(z)}[3(1-\Omega_{m_0})(1-\omega_\sigma)(1+z)^{3(1+\omega_\sigma)-2}-\frac{1}{2}\alpha\Omega_{m_0}e^{-\frac{1}{2}\alpha\sigma}(1+z)^2\frac{d\sigma}{dz}]^{\frac{1}{2}}\end{equation}
where
$\chi(z)=[(1-\Omega_{m_0})(1+z)^{3(1+\omega_\sigma)}+\Omega_{m_0}e^{-\frac{1}{2}\alpha\sigma}(1+z)^3]^{\frac{1}{2}}$.
According Eqs.(19)(20), we can construct the DCQ potential directly.
In fact the constant $\alpha$ reflects the coupled intensity. When
the coupled constant $\alpha\rightarrow0$, DCQ model will reduce to
quintessence model and Eqs.(19)(20) become respective
\begin{equation}\tilde{W}(z)=\frac{1}{2}(1-\Omega_{m_0})(1-\omega_\sigma)(1+z)^{3(1+\omega_\sigma)}\end{equation}
\begin{equation}\frac{d\sigma}{dz}=-\frac{1}{\chi(z)}[3(1-\Omega_{m_0})(1-\omega_\sigma)(1+z)^{3(1+\omega_\sigma)-2}]^{\frac{1}{2}}\end{equation}
\vskip 0.3 in
\begin{flushleft}\textbf{3. Specific Cases}\end{flushleft}
Now let us consider four cases[26-31]of the $\omega_\sigma(z)$ in
DCQ model, which fit the observations well. \vskip 0.15 in
\par \textbf{Case I}: $\omega_\sigma=\omega_0$[26]
\begin{equation}\tilde{W}(z)=\frac{1}{2}(1-\Omega_{m_0})(1-\omega_0)(1+z)^{3(1+\omega_0)}+\frac{1}{12}\alpha\Omega_{m_0}e^{-\frac{1}{2}\alpha\sigma}(1+z)^4\frac{d\sigma}{dz}\end{equation}
\begin{equation}\frac{d\sigma}{dz}=-\frac{1}{\chi(z)}[3(1-\Omega_{m_0})(1-\omega_0)(1+z)^{3(1+\omega_0)-2}-\frac{1}{2}\alpha\Omega_{m_0}e^{-\frac{1}{2}\alpha\sigma}(1+z)^2\frac{d\sigma}{dz}]^{\frac{1}{2}}\end{equation}
\vskip 0.15 in
\par \textbf{Case II}: $\omega_\sigma=\omega_0+\omega_1z$[27]
\begin{equation}\tilde{W}(z)=\frac{1}{2}(1-\Omega_{m_0})(1-\omega_0-\omega_1z)(1+z)^{3(1+\omega_0+\omega_1z)}+\frac{1}{12}\alpha\Omega_{m_0}e^{-\frac{1}{2}\alpha\sigma}(1+z)^4\frac{d\sigma}{dz}\end{equation}
\begin{equation}\frac{d\sigma}{dz}=-\frac{1}{\chi(z)}[3(1-\Omega_{m_0})(1-\omega_0-\omega_1z)(1+z)^{3(1+\omega_0+\omega_1z)-2}-\frac{1}{2}\alpha\Omega_{m_0}e^{-\frac{1}{2}\alpha\sigma}(1+z)^2\frac{d\sigma}{dz}]^{\frac{1}{2}}\end{equation}
\vskip 0.15 in
\par \textbf{Case III}: $\omega_\sigma=\omega_0+\omega_1\frac{z}{1+z}$[28-30]
\begin{equation}\tilde{W}(z)=\frac{1}{2}(1-\Omega_{m_0})(1-\omega_0-\omega_1\frac{z}{1+z})(1+z)^{3(1+\omega_0+\omega_1\frac{z}{1+z})}+\frac{1}{12}\alpha\Omega_{m_0}e^{-\frac{1}{2}\alpha\sigma}(1+z)^4\frac{d\sigma}{dz}\end{equation}
\begin{equation}\frac{d\sigma}{dz}=-\frac{1}{\chi(z)}[3(1-\Omega_{m_0})(1-\omega_0-\omega_1\frac{z}{1+z})(1+z)^{3(1+\omega_0+\omega_1\frac{z}{1+z})-2}-\frac{1}{2}\alpha\Omega_{m_0}e^{-\frac{1}{2}\alpha\sigma}(1+z)^2\frac{d\sigma}{dz}]^{\frac{1}{2}}\end{equation}
\par \textbf{Case IV}: $\omega_\sigma=\omega_0+\omega_1ln(1+z)$[31]
\begin{equation}\tilde{W}(z)=\frac{1}{2}(1-\Omega_{m_0})[1-\omega_0-\omega_1ln(1+z)](1+z)^{3(1+\omega_0+\omega_1ln(1+z))}+\frac{1}{12}\alpha\Omega_{m_0}e^{-\frac{1}{2}\alpha\sigma}(1+z)^4\frac{d\sigma}{dz}\end{equation}
\begin{equation}\frac{d\sigma}{dz}=-\frac{1}{\chi(z)}\{3(1-\Omega_{m_0})[1-\omega_0-\omega_1ln(1+z)](1+z)^{3(1+\omega_0+\omega_1ln(1+z))-2}-\frac{1}{2}\alpha\Omega_{m_0}e^{-\frac{1}{2}\alpha\sigma}(1+z)^2\frac{d\sigma}{dz}\}^{\frac{1}{2}}\end{equation}

\vskip 0.3 in
\begin{minipage}{0.5\textwidth}
\includegraphics[scale=0.8]{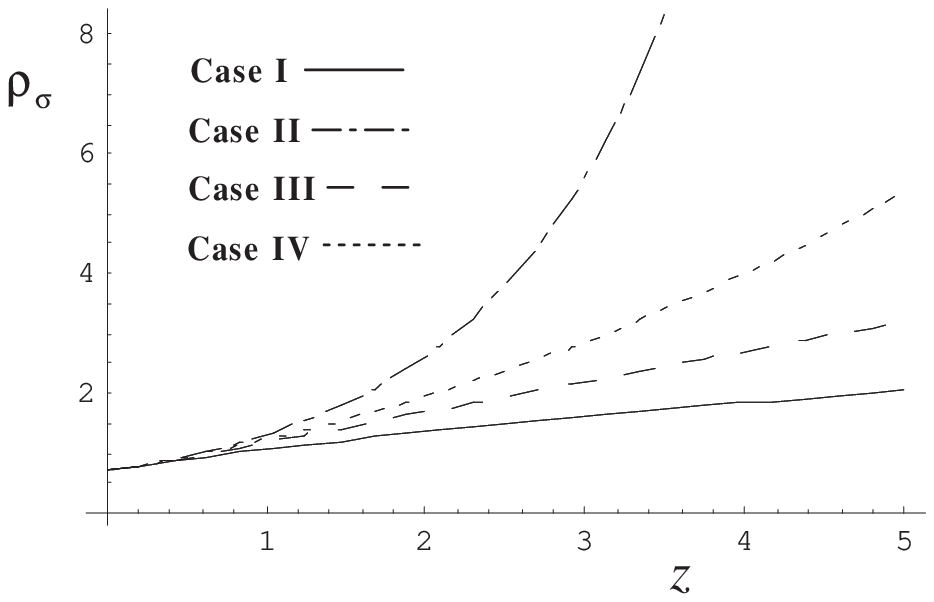}
{\small Fig.1 The evolution of the dark energy density $\rho_\sigma$
with respect to $z$ in the four cases in DCQ model. We set
$\alpha=0.0005$, $\Omega_{m_0}=0.3$, $\omega_0=-0.8$, $\omega_1=0.1$
and $\sigma_0=0.8$.}
\end{minipage}
\hspace{0.02\textwidth}
\begin{minipage}{0.5\textwidth}
\includegraphics[scale=0.8]{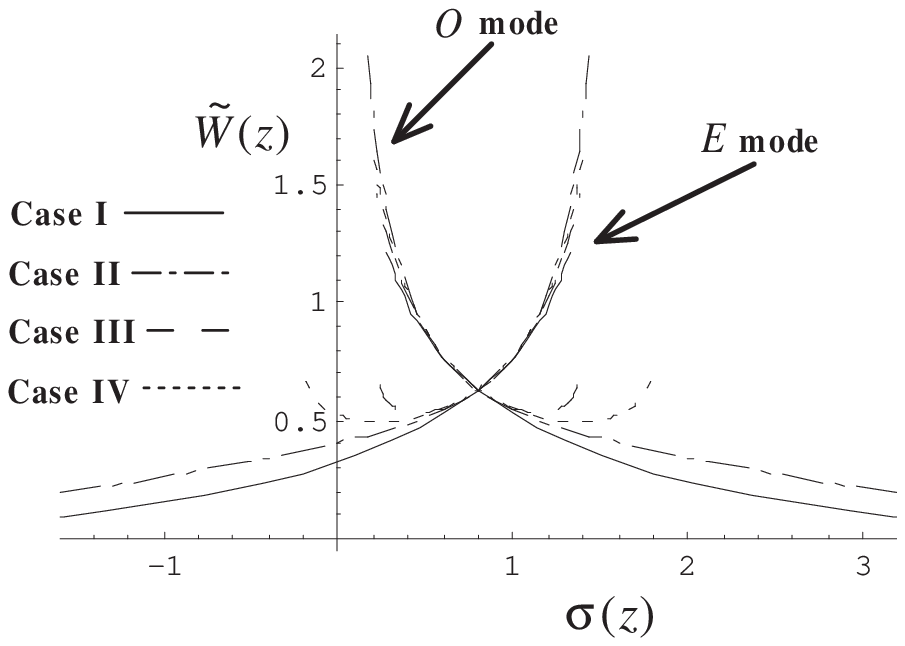}
{\small Fig.2 Constructed DCQ potential in the four cases. We set
$\alpha=0.0005$, $\Omega_{m_0}=0.3$, $\omega_0=-0.8$, $\omega_1=0.1$
and $\sigma_0=0.8$.}
\end{minipage}
\hspace{0.02\textwidth}
\begin{minipage}{0.49\textwidth}
\includegraphics[scale=0.8]{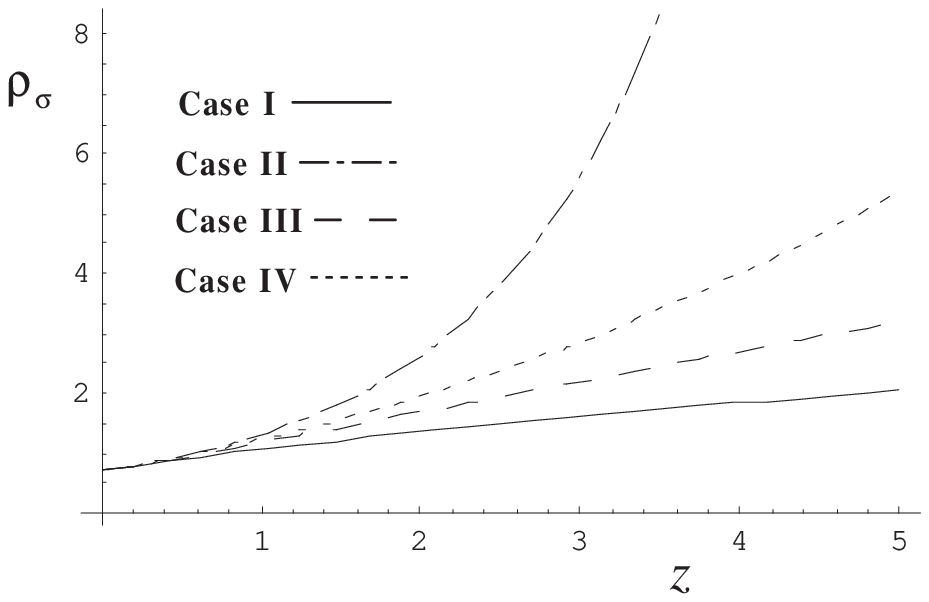}
{\small Fig.3 The evolution of the dark energy density $\rho_\sigma$
with respect to $z$ in the four cases when the coupling between
dilaton and matter is zero($\alpha=0$).}
\end{minipage}
\hspace{0.02\textwidth}
\begin{minipage}{0.5\textwidth}
\includegraphics[scale=0.8]{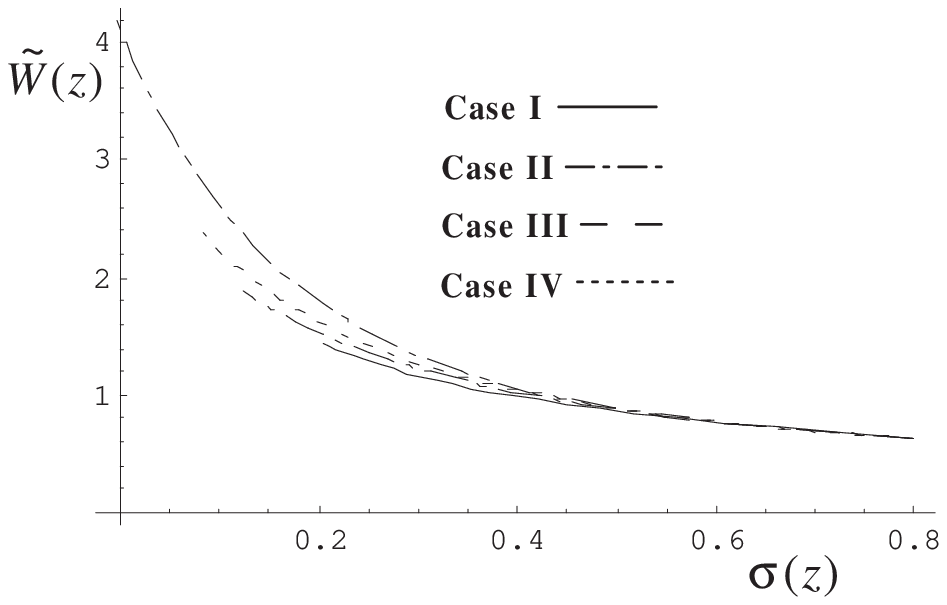}
{\small Fig.4 Quintessence potential in the four cases when the
coupling between dilaton and matter is zero($\alpha=0$).}
\end{minipage}

\vskip 0.3 in
\par Fig.1 shows the evolution of the dark energy
density $\rho_\sigma$ with respect to $z$ in the four cases in DCQ
model. We can see that the slope of $\rho_\sigma$ with respect to
$z$ becomes more steeper and more from Case I, Case III, Case IV to
Case II. However, $\rho_\sigma$ in the four cases tends to be the
same evolutive behavior in range of $0<z<1$($0.422<\sigma<0.8$)
while they differ when the redshift $z$ becomes large. The evolution
of the constructed DCQ potential $\tilde{W}(z)$ with respect to the
DCQ field is shown in Fig.2. We can easily see that the shape of
$\tilde{W}(z)$ is uniform in the range of
$0.422<\sigma<0.8$($0<z<1$) while they differ beyond this range. To
show difference between the DCQ model and the quintessence model, we
also plot the evolution of $\rho_\sigma$ and the constructed
potential when coupling term becomes zero($\alpha=0$) in Figs.3-4.
Comprising Fig.1 with Fig.3, we know the evolution of $\rho_\sigma$
with respect to $z$ is hardly distinct between the DCQ model and the
quintessence model. However the shape of DCQ potentials are quite
different from those of the quintessence potentials as shown in
Fig.2 and Fig.4. In Fig.4, quintessence potentials are all in the
form of runaway type[24], which is related to the supersymmetric
theories and tachyons in superstring theories interestingly.
However, in DCQ model, the evolution of DCQ potential with respect
to DCQ field is divided into two different modes called $"O"$ mode
and $"E"$ mode, because of the existence of coupling term. We see
that the $"O"$ mode possesses evolutive behavior of runaway type,
just like quintessence potentials. On the contrary, DCQ potential in
$"E"$ mode tends to be infinite when $\sigma\rightarrow\infty$,
which results in an unstable dynamic system.

\vskip 0.5 in
\begin{flushleft}\textbf{{3. Conclusions}}\end{flushleft}
In this paper, we apply the method of constructing quintessence
potential to the DCQ model. Based on this, we investigate the
evolutive behavior of the dark energy density $\rho_\sigma$ with
respect to $z$ and the constructed DCQ potential with respect to
$\sigma$ in four cases of the equation of state $\omega(z)$, which
fit the observations well. These results are shown mathematically in
the plots. According to the comparision between the constructed DCQ
potential and quintessence potential($\alpha=0$), we find that the
shapes of the constructed DCQ potential quite different from the one
of quintessence potential. DCQ potentials possess two different
evolutive mode $"O"$ and $"E"$. DCQ potentials in the $"E"$ mode
will lead to an unstable dynamic system, so, $"E"$ mode will be
ruled out. DCQ potential in $"O"$ mode belongs to the form of
runaway type, which is related to the supersymmetric theories and
tachyons in superstring theories interestingly[36].

\begin{flushleft}\textbf{Acknowledgements}\end{flushleft}
This work is partially supported by National Nature Science
Foundation of China under Grant No.10573012 and HHIT scientific
research startup fund under Grant No.KK06033.

\begin{flushleft}{\noindent\bf References}
 \small{
\item {1.}{ A. G. Riess et al., \textit{Astrophys. J}\textbf{607}, 665(2004);
\\\hspace{0.15 in}A. G. Riess, \textit{Astron. J}\textbf{116}, 1009(1998);
\\\hspace{0.15 in}S. Perlmutter et al., \textit{Astrophys. J}\textbf{517}, 565(1999);
\\\hspace{0.15 in}N. A. Bahcall et al., \textit{Science}\textbf{284}, 1481(1999).}
\item {2.}{ D. N. Spergel et al., Astrophys. J. Suppl\textbf{148}, 175(2003).}
\item {3.}{ P. de Bernardis et al., astro-ph/0105296;
\\\hspace{0.15 in}R. Stompor et al., astro-ph/0105062;
\\\hspace{0.15 in}C. Pryke et al., astro-ph/0104490.}
\item {4.}{ J. S. Bagla, H. K. Jassal and T. Padmamabhan, \textit{Phys. Rev. D}\textbf{67}, 063504(2003).}
\item {5.}{ L. Amendola, M. Quartin, S. Tsujikawa and I. Waga, \textit{Phys.Rev.D}\textbf{74}, 023525(2006);
\\\hspace{0.15 in}E. Elizalde, S. Nojiri and S. D. Odintsov, arXiv:hep-th/0405034;
\\\hspace{0.15 in}S. Nojiri, S. D. Odintsov and M. Sasaki, \textit{Phys.Rev. D}\textbf{70} 043539(2004);
\\\hspace{0.15 in}S. Nojiri and S. D. Odintsov, arXiv:hep-th/0601213;
\\\hspace{0.15 in}S. Nojiri and S. D. Odintsov \textit{Phys. Lett. B}\textbf{639}, 144(2006);
\\\hspace{0.15 in}S. Nojiri, S. D. Odintsov and H. Stefancic, arXiv:hep-th/0608168;
\\\hspace{0.15 in}S. Nojiri and S. D. Odintsov, \textit{Phys. Lett. B}\textbf{562}, 147(2003)[arXiv:hep-th/0303117];
\\\hspace{0.15 in}S. Nojiri and S.D. Odintsov, \textit{Phys. Rev. D}\textbf{70}, 103522(2004)[arXiv:hep-th 0408170];
\\\hspace{0.15 in}S. Nojiri, S. D. Odintsov and S. Tsujikawa, arXiv:hep-th/0501025;
\\\hspace{0.15 in}S. Nojiri and S. D. Odintsov, \textit{Phys. Rev. D}\textbf{7}2, 023003(2005)[arXiv:hep-th/0505215];
\\\hspace{0.15 in}S. Nojiri and S. D. Odintsov, arXiv:hep-th 0506212;
\\\hspace{0.15 in}S. Nojiri and S. D. Odintsov, arXiv:hep-th/0611071;
\\\hspace{0.15 in}B. Boisseau et al., \textit{Phys. Rev. Lett}\textbf{85}, 2236(2000)[arXiv:gr-qc/0001066];
\\\hspace{0.15 in}G. Esposito-Farese and D. Polarski, \textit{Phys. Rev. D}\textbf{63}, 063504(2001)[arXiv:gr-qc/0009034];
\\\hspace{0.15 in}Xin Zhang, \textit{Mod. Phys. Lett. A}\textbf{20}, 2575(2005)[arXiv:astro-ph/0503072];
\\\hspace{0.15 in}Xin Zhang, \textit{Phys. Lett. B}\textbf{611}, 1(2005)[arXiv:astro-ph/0503075];
\\\hspace{0.15 in}M. R. Setare, \textit{Phys. Lett. B}\textbf{642},1(2006)[arXiv:hep-th/0609069];
\\\hspace{0.15 in}M. R  Setare, arXiv:hep-th/0609104;
\\\hspace{0.15 in}M. R. Setare, arXiv:hep-th/0610190.}
\item {6.}{ C. Wetterich \textit{Nucl. Phys. B}\textbf{302}, 668(1998);
\\\hspace{0.15 in}E. J. Copeland, M. Sami and S. Tsujikawa, arXiv:hep-th/060305;
\\\hspace{0.15 in}P. G. Ferreira and M. Joyce \textit{Phys. Rev. D}\textbf{58}, 023503(1998);
\\\hspace{0.15 in}J. Frieman, C. T. Hill, A. Stebbinsand and I.Waga, \textit{Phys. Rev. Lett}\textbf{75}, 2077(1995);
\\\hspace{0.15 in}P. Brax and J. Martin, \textit{Phys. Rev. D}\textbf{61}, 103502(2000)
\\\hspace{0.15 in}T. Barreiro, E. J. Copeland and N. J. Nunes, \textit{Phys. Rev. D}\textbf{61}, 127301(2000);
\\\hspace{0.15 in}I. Zlatev, L. Wang and P. J. Steinhardt \textit{Phys. Rev. Lett} \textbf{82}, 896(1999).}
\item {7.}{ T. Padmanabhan, and T. R. Choudhury, \textit{Phys. Rev. D}\textbf{66}, 081301(2002).}
\item {8.}{ A. Sen, \textit{JHEP} \textbf{0204}, 048(2002).}
\item {9.}{ C. Armendariz-Picon, T. Damour and V. Mukhanov, \textit{Phys. Lett. B}\textbf{458}, 209(1999).}
\item {10.}{A. Feinstein, \textit{Phys. Rev. D}\textbf{66}, 063511(2002).}
\item {11.}{M. Fairbairn and M. H. Tytgat, \textit{Phys. Lett. B}\textbf{546}, 1(2002).}
\item {12.}{A. Frolov, L. Kofman and A. Starobinsky, \textit{Phys.Lett.B} \textbf{545}, 8(2002).}
\item {13.}{C. Acatrinei and C. Sochichiu, \textit{Mod. Phys. Lett. A}\textbf{18}, 31(2003);
\\\hspace{0.17 in}S. H. Alexander, \textit{Phys. Rev. D}\textbf{65}, 0203507(2002).}
\item{14.}{A. Mazumadar, S. Panda and A. Perez-Lorenzana, \textit{Nucl. Phys. B}\textbf{614}, 101(2001);
\\\hspace{0.17 in}S. Sarangi and S. H. Tye,
\textit{Phys. Lett. B}\textbf{536}, 185(2002).}
\item{15.}{C. Armend\'{a}riz-Pic\'{o}n, V. Mukhanov and P. J. Steinhardt, \textit{Phys.Rev.Lett}\textbf{85}, 4438(2000);
\\\hspace{0.17 in}C. Armend\'{a}riz-Pic\'{o}n, V. Mukhanov and P. J. Steinhardt, \textit{Phys. Rev. D}\textbf{63}, 103510(2001);
\\\hspace{0.17 in}T. Chiba, \textit{Phys.Rev.D}\textbf{66}, 063514(2002);
\\\hspace{0.17 in}M. Malquarti, E. J. Copeland, A. R. Liddle and M. Trodden, \textit{Phys. Rev. D}\textbf{67}, 123503(2003);
\\\hspace{0.17 in}R. J. Sherrer, \textit{Phys. Rev. Lett}\textbf{93)}, 011301(2004);
\\\hspace{0.17 in}L. P. Chimento, \textit{Phys. Rev. D}\textbf{69}, 123517(2004);
\\\hspace{0.17 in}A. Melchiorri, L. Mersini, C. J. Odman and M. Trodden, \textit{Phys. Rev. D}\textbf{68}, 043509(2003).}
\item{16.}{A. Sen, \textit{JHEP} \textbf{0207}, 065(2002);
\\\hspace{0.17 in}M. R. Garousi, \textit{Nucl. Phys. B}\textbf{584}, 284(2000);
\\\hspace{0.17 in}M. R. Garousi, \textit{JHEP} \textbf{0305}, 058(2003);
\\\hspace{0.17 in}E. A. Bergshoeff, M. de Roo, T. C. de Wit, E. Eyras and S. Panda,\textit{ JHEP} \textbf{0005}, 009(2000);
\\\hspace{0.17 in}J. Kluson, \textit{Phys. Rev. D}\textbf{62}, 126003(2000);
\\\hspace{0.17 in}G. W. Gibbons, \textit{Phys. Lett. B}\textbf{537}, 1(2002);
\\\hspace{0.17 in}M. Sami, P. Chingangbam and T. Qureshi, \textit{Phys. Rev. D}\textbf{66}, 043530(2002);
\\\hspace{0.17 in}M. Sami, \textit{Mod. Phys. Lett. A}\textbf{18}, 691(2003);
\\\hspace{0.17 in}Y. S. Piao, R. G. Cai, X. m. Zhang and Y. Z. Zhang, \textit{Phys. Rev. D}\textbf{66},121301(2002);
\\\hspace{0.17 in}L. Kofman and A. Linde, \textit{JHEP} \textbf{0207}, 004(2002).}
\item{17.}{H. Q. Lu, \textit{Int. J. Mod. Phys. D}\textbf{14}, 355(2005);
\\\hspace{0.17 in}W. Fang, H. Q. Lu, Z. G. Huang and K. F. Zhang, \textit{Int. J. Mod. Phys. D}\textbf{15}, 199(2006)[arXiv:hep-th/0409080];
\\\hspace{0.17 in}W. Fang, H. Q. Lu and Z. G. Huang, arXiv:hep-th/0606032;
\\\hspace{0.17 in}Z. G. Huang, X. H. Li and Q. Q. Sun, arXiv:hep-th/0610019.}
\item{18.}{X. Z. Li and J. G. Hao, \textit{Phys. Rev. D}\textbf{69}, 107303(2004).}
\item{19.}{T. Chiba, T. Okabe and M. Yamaguchi, \textit{Phys. Rev. D}\textbf{62}, 023511(2000).
\\\hspace{0.17 in}L. Amendola, S. Tsujikawa, and M. Sami, \textit{Phys. Lett. B}\textbf{632}, 155(2006);
\\\hspace{0.17 in}L. Amendola, \textit{Phys. Rev. Lett.}\textbf{93}, 181102(2004).}
\item{20.}{P. Singh, M. Sami and N. Dadhich, \textit{Phys.Rev. D}\textbf{68}, 023522(2003);
\\\hspace{0.17 in}S. M. Carroll, M. Hoddman and M. Trodden, \textit{Phys. Rev. D}\textbf{68}, 023509(2003).}
\item{21.}{W. Hao, R. G. Cai and D. F. Zeng, \textit{Class.Quant.Grav}\textbf{22}, 3189(2005);
\\\hspace{0.17 in}Z. K. Guo, Y. S. Piao, X. M. Zhang, Y.Z. Zhang, \textit{Phys.Lett. B}\textbf{608}, 177(2005);
\\\hspace{0.17 in}B. Feng, arXiv:astro-ph/0602156.}
\item{22.}{Q. G. Huang and M. Li, \textit{JCAP}\textbf{0408}, 013(2004);
\\\hspace{0.17 in}M. Ito, \textit{Europhys. Lett}.\textbf{71}, 712-715(2005);
\\\hspace{0.17 in}K. Ke and M. Li, \textit{Phys.Lett.B}\textbf{606}, 173-176(2005);
\\\hspace{0.17 in}Q. G. Huang and M. Li, \textit{JCAP}\textbf{0503}, 001(2005);
\\\hspace{0.17 in}Y. G. Gong, B. Wang and Y. Z. Zhang, \textit{Phys. Rev. D}\textbf{72}, 043510(2005);
\\\hspace{0.17 in}X. Zhang, \textit{Int. J. Mod. Phys. D}\textbf{14}, 1597-1606(2005).}
\item{23.}{D. Huterer and M. S. Turner,\textit{ Phys. Rev. D}\textbf{60}, 081301(1999);
\\\hspace{0.17 in}A. A. Starobinsky, \textit{JETP Lett.}\textbf{68}, 757(1998);
\\\hspace{0.17 in}T. Chiba and T. Nakamura, \textit{Phys. Rev. D}\textbf{62}, 121301(2000).}
\item{24.}{Z. K. Guo, N. Ohta and Y. Z. Zhang, \textit{Phys. Rev. D}\textbf{72}, 023504(2005);
\\\hspace{0.17 in}Z. K. Guo, N. Ohta and Y. Z. Zhang, \textit{Mod. Phys. Lett. A}\textbf{22}, 883(2007)[arXiv:astro-ph/0603109].}
\item{25.}{Z. G. Huang, H. Q. Lu and W. Fang, \textit{Class. Quant. Grav.}\textbf{23}, 6215(2006)[arXiv:hep-th/0604160];
\\\hspace{0.17 in}Z. G. Huang, H. Q. Lu and W. Fang, \textit{Int. J. Mod. Phys. D.}\textbf{15}, 1501(2006);
\\\hspace{0.17 in}Z. G. Huang, H. Q. Lu, W. Fang and K. F. Zhang, \textit{Astrophys. Space Sci.}\textbf{305}, 177(2006);
\\\hspace{0.17 in}Z. G. Huang, H. Q. Lu and W. Fang, arXiv:hep-th/0610018.}
\item{26.}{S. Hannestad and E. Mortsell, \textit{Phys. Rev. D}\textbf{66}, 063508(2002).}
\item{27.}{A. R. Cooray and D. Huterer, \textit{Astrophys. J.}\textbf{513}, L95(1999).}
\item{28.}{M. Chevallier and D. Polarski, \textit{Int. J. Mod. Phys. D}\textbf{10}, 213(2001). }
\item{29.}{E. V. Linder, \textit{Phys. Rev. Lett.}\textbf{90}, 091301(2003).}
\item{30.}{T. Padmanabhan and T.R. Choudhury, \textit{Mon. Not. Roy. Astron. Soc.}\textbf{344}, 823(2003).}
\item{31.}{B. F. Gerke and G. Efstathiou, \textit{Mon. Not. Roy. Astron. Soc.}\textbf{335}, 33(2002).}
\item{32.}{C. M. Will, \textit{Living Rev. Rel.}\textbf{4}, 4(2001).}
\item{33.}{Y. M. Cho, \textit{Phys. Rev.Lett}\textbf{68}, 3133(1992).}
\item{34.}{T. Damour and K. Nordtvedt, \textit{Phys. Rev. Lett}\textbf{70}, 2217(1993).}
\item{35.}{B. Bertotti, L. Iess and P. Tortora, \textit{Nature}\textbf{425}, 374(2003).}
\item{36.}{A. Sen, \textit{Phys. Scripta}\textbf{T117}, 70(2005).}

}
\end{flushleft}
\end{document}